\documentstyle[aps,multicol,epsfig]{revtex}
\newcommand \be {\begin{equation}}
\newcommand \ee {\end{equation}}
\begin{document}

\draft

\title{An entropy--driven noise--induced phase transition}
\author{M. Iba\~nes$^{1}$ , J. Garc\'{\i}a-Ojalvo$^{2}$, R. Toral$^3$
and J.M. Sancho$^{1}$}

\address{$^{1}$
Department d'Estructura i Constituents de la Mat\`eria,
Universitat de Barcelona, Diagonal 647, E--08028 Barcelona, Spain\\
$^2$ Departament de F\'{\i}sica i Enginyeria Nuclear, Universitat
Polit\`ecnica de Catalunya, Colom 11, E--08222 Terrassa, Spain\\
$^3$
Instituto Mediterr\'aneo de Estudios Avanzados (CSIC-UIB),
E-07071 Palma de Mallorca, Spain
}

\date{\today }
\maketitle

\begin{abstract}
We introduce a class of exactly solvable models which exhibit an
ordering noise-induced phase transition driven by an entropic mechanism.
In contrast with previous studies, order does not appear in this
case as a result of an instability of the disordered phase produced by noise,
but of a balance between the relaxing deterministic dynamics and the
randomizing (entropic) character of the fluctuations.
A finite-size scaling analysis of the phase transition reveals that
it belongs to the universality class of the equilibrium Ising model.
All these results are analyzed in the light of the nonequilibrium
probability distribution of the system, which can be derived
analytically. The relevance of our result as a possible scenario of the
so-called Lower Critical Solution Temperature (LCST) transitions
is analysed.

\vspace{8pt}
\vspace{8pt}
\end{abstract}

\begin{multicols}{2}

\vskip1cm 

A surprising discovery in the field of stochastic processes is that the
interaction between noise and spatial degrees of freedom in extended nonlinear
systems can lead, under appropriate conditions, to noise-induced order
\cite{nises}. This seemingly counterintuitive phenomenon arises when a
distributed medium develops some kind of spatiotemporal self-organization which
requires noise to exist. The resulting ordered state can take many forms, such
as a static spatial pattern \cite{prl}, traveling excitable waves \cite{jung},
propagating pulses \cite{epl}, spatiotemporal chaotic states \cite{spiral}, or
a coherent response to an external periodic forcing \cite{alexei}, among
others. From a fundamental point of view, however, the most interesting example
is that of noise--induced ordering phase transitions (NIOPTs) \cite{broeck}. In
this case, transitions between true extended phases (in a thermodynamic sense)
are produced when the noise intensity is used as control parameter. These
non-equilibrium phase transitions can therefore be characterized using standard
techniques of critical phenomena, such as dynamic renormalization group,
mean-field approximations, as well as finite-size scaling analyses using, for
instance, results obtained from numerical simulations of stochastic partial
differential equations. Up to now, all the arguments presented to account for
the occurrence of NIOPTs have been dynamical ones. This is mainly because
it has been impossible so far to find a model whose nonequilibrium steady-state
probability distribution and the associated effective potential are known. In
particular, noise-induced phase transitions have been systematically explained
in terms of a short-time instability of the local dynamics, which becomes
``frozen'' at larger times by the spatial coupling \cite{broeck,tutorial}. 

In this Letter we introduce a class of systems exhibiting NIOPTs for which the
steady-state probability distribution can be obtained {\em exactly}, so that
one can define the corresponding nonequilibrium free energy or potential. As a
consequence, the NIOPT can be studied in the steady state, with no dynamical
reference. It turns out that the noise--induced phase transition that appears
in these model systems is not a consequence of an instability of the disordered
homogeneous state. Rather, the situation is similar to what happens for
noise--induced transitions in 0--dimensional systems, where the disordered
state is linearly stable, but the effective non-equilibrium potential in the
steady state is bimodal \cite{HLNIT,MT97}. A
phase transition is obtained by coupling adequately many of those
0-dimensional systems. For a strong enough spatial coupling, the neighboring
variables tend to a common value and a macroscopic ordered phase appears as
a result of ergodicity breaking. This
situation contrasts vividly with NIOPTs reported so far, where the need of a
short-time dynamical instability prevents systems that undergo noise-induced
transitions in 0-d from exhibiting NIOPTs in the presence of spatial coupling
\cite{broeck}.

The phase transition that we report in what follows shares some features
with the so-called Lower Critical Solution Temperature (LCST)
transition which appears in some polymer blends \cite{snyder}. In these very
special systems, one can go from a disordered phase (homogeneous mixture) to an
ordered phase (separated phases) by increasing the temperature (inverted phase
diagram). This is an equilibrium  phenomenology which goes in the opposite
direction of what is expected intuitively from statistical physics. Our
findings present a possible scenario, in terms of a Langevin description, in
which these transitions can take place. To reach our goal we consider two
important theoretical ingredients: the presence of a field-dependent kinetic
coefficient and the existence of a fluctuation-dissipation relation
accordingly. 

Let us consider the following generic
deterministic model for a real field $\phi(\vec r,t)$:
\be
\label{eq:1a}
\frac{\partial \phi(\vec r,t)}{\partial t} = -\Gamma(\phi(\vec r,t)) \frac{\delta V}{\delta
\phi(\vec r,t)},
\ee
which corresponds to a relaxational flow \cite{mhsm} in a potential
$V(\{\phi\})$ with a field-dependent kinetic coefficient $\Gamma(\phi)$.
Field-dependent
kinetic coefficients can appear in coarse-grained, master-equation
derivations of macroscopic (deterministic or stochastic) field equations
\cite{langer,kitahara1}. Their explicit form depends on the assumptions in
their derivation, but the general trend is that the coefficient is large in the
disordered phase and small in the ordered phase \cite{kitahara1,martin}.
These coefficients have been introduced in the
context of phase separation in binary mixtures, either to explore ad hoc their
influence, or by necessity to take into account the effect of a gravitational
field \cite{kitahara2}. In those cases, the kinetic
coefficients are also a function of the spatial derivatives of the field.
in order to reflect the conserved nature of the concentration field.
Here, on the other hand, we consider for simplicity a non-conserved situation.

When stochastic terms are introduced in this description a
fluctuation-dissipation relation is used \cite{kitahara1}. Following this
approach we define that the Langevin equation (in the Stratonovich
interpretation) corresponding to the deterministic equation (\ref{eq:1a})
to be:
\be
\label{eq:1}
\frac{\partial \phi(\vec r,t)}{\partial t} = -\Gamma(\phi(\vec r,t)) \frac{\delta V}{\delta
\phi(\vec r,t)} +
\Gamma(\phi(\vec r,t))^{1/2}\xi(\vec r,t)\,.
\ee 
The noise is taken to be Gaussian, with zero mean
and correlation 
\be
\langle \xi(\vec r,t) \xi(\vec r\,',t') \rangle =
2 \varepsilon \,\delta(\vec r-\vec r\,') \delta(t-t')\,,
\ee
where $\varepsilon$ is the noise intensity.
Under these conditions, the stationary solution of the probability distribution for the field, $P_{\rm st}(\{\phi\})$ is of the Boltzmann's type:
\be
\label{pst}
P_{st}(\{\phi \}) \sim {\rm e}^{- V_{\rm eff}/\varepsilon}\,,
\ee
in terms of an effective potential
\be
V_{\rm eff}(\{\phi\}) \equiv V(\{\phi\}) + \frac{\varepsilon_0}{2}
\int \, d\vec r \ln \Gamma(\phi(\vec r))\, ,
\label{freeenergy}
\ee
where $\varepsilon_0$ is  a renormalized parameter, proportional to
$\varepsilon$, which  includes
an ultraviolet cutoff \cite{cutoff}.
The previous expression defines the exact nonequilibrium free energy   
of model (\ref{eq:1}). We will now show that this class of systems
can exhibit a noise-induced phase transition.

\begin{figure}
\narrowtext
\centerline{\epsfig{file=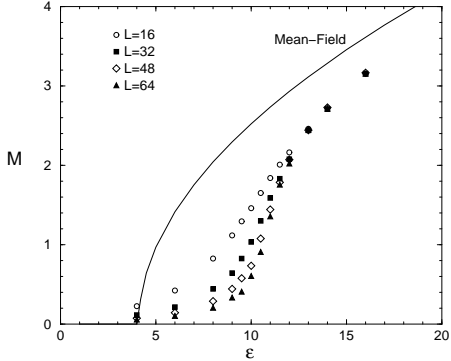,width=6cm}}
\caption{\label{fig:1} Intensive order parameter $M$ as function of the
noise intensity for $a=1$, $c=0.5$, $K=1$ and for four different system
sizes. The continuous line is the Weiss mean--field result. 
}
\end{figure}

\begin{figure}
\narrowtext
\centerline{\epsfig{file=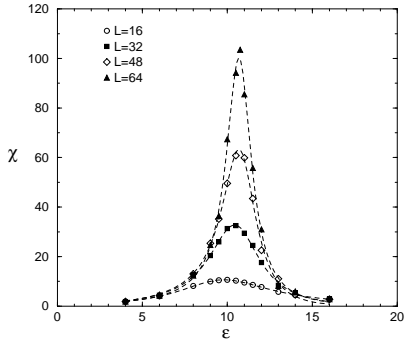,width=6cm}}
\caption{\label{fig:2} Generalized susceptibility $\chi$ versus
noise intensity for the case of Fig. \protect\ref{fig:1}.
}
\end{figure}

So far, the discussion has been made in very general terms. In order to
proceed further, we assume that the deterministic free-energy potential
has the usual Landau form 
\be
V(\{\phi\}) = \int\,d\vec r\left [V_0(\phi(\vec r)) + 
\frac{K}{2}|\vec \nabla \phi(\vec r)| ^2\right]\,,
\ee
and we choose a monostable local potential
$V_0(\phi) = \frac{a}{2}\phi^2$, with $a>0$.
We also consider a non-conserved order parameter system in which the kinetic
coefficient $\Gamma$ is only a function of the field and not of its spatial
derivatives. According to the discussion above, we adopt the functional form
\be
\label{eq:M}
\Gamma(\phi)=1/(1+c\phi^2)\,,
\ee
which means that fluctuations are larger in dilute regions and smaller
in dense ones. In this particular case, and
according to Eqs. (\ref{pst}) and (\ref{freeenergy}), the {\em local}
effective potential can be seen to become bistable for $\varepsilon_0>
a/c$. One can then expect that for a strong enough coupling $K$, a true
{\em phase} transition towards an ordered state controlled by the noise
intensity $\varepsilon$ exists. In order to check this
expectation, we perform a Weiss mean-field analysis of the corresponding
local effective potential \cite{bpah}. The analysis is more easily done
by considering a
spatial discretization in a regular d--dimensional lattice with spacing
$\Delta x=1$, i.e. $\phi(\vec r_i)\to\phi_i$, with $i$ the cell index. The
gradient term is then approximated by the sum over
nearest neighbors on the lattice in the standard way,
$|\vec \nabla\phi|^2\to\sum_{j=1}^{d}(\phi_j-\phi_i)^2$.
The mean-field approximation
is tantamount to replacing the exact value of the neighbors by a mean-field
common value $M$. In this way, the local effective potential becomes
\be
V_{\rm eff}(\phi,M) = V_0(\phi) + \frac{\varepsilon_0}{2} \ln \Gamma(\phi)
+ 
d\,K(\phi-M)^2\,,
\ee
and depends on the yet unknown mean--field value $M$. This can be 
computed by using the consistency relation:
\be
M=\frac{\int_{-\infty}^{\infty} \phi\,
{\rm e}^{-V_{\rm eff}(\phi,M)/\varepsilon}
d\phi}{\int_{-\infty}^{\infty}
{\rm e}^{-V_{\rm eff}(\phi,M)/\varepsilon} d\phi}\,,
\ee
which, in general, has to be solved numerically. The solution of this
equation is
shown as a solid line in Fig. \ref{fig:1} for the two-dimensional case.
The result predicts a 
phase transition from disorder ($M=0$) to order ($M\neq 0$) as the noise
intensity increases.
This theoretical result has been confirmed by numerical
simulations of a discretized version of model (\ref{eq:1}) in a
two-dimensional square lattice.
Following standard procedures of equilibrium statistical mechanics,
we have characterized the phase transition by means of two quantities:
the {\em intensive order parameter} $M=\langle m \rangle$, where
$m=|\sum_i \phi_i|/L^2$ is the absolute value of the spatial average
of the field, and the {\em generalized susceptibility} 
\be
\chi=L^2\,\frac{\langle m^2 \rangle-\langle m \rangle^2}{\varepsilon}\,,
\ee
which measures the relative fluctuations of the field in its steady state  
($\varepsilon$ plays the role of $K_{B}T$ in  
equilibrium phase transitions and $L$ is the linear system size).
The numerical results are plotted in Figs. \ref{fig:1} and \ref{fig:2},
which show the behavior of the order parameter $M$ and the susceptibility
$\chi$, respectively, as the noise intensity $\varepsilon$ increases,
for four different system sizes. The figures clearly denote the existence of 
a noise--induced ordering phase transition with the usual features of a
second--order
equilibrium phase transition, in agreement with the theoretical
analysis. Monte Carlo simulations of the system making use of the
nonequilibrium effective free energy give completely compatible results.
The position of the critical point has been computed both by extrapolating
to infinite size the location of the maxima of the susceptibility and by
means of a finite-size scaling analysis (see below). The two approaches
lead to similar results.

As mentioned above, noise--induced phase transitions have been related
so far to an instability of the homogeneous phase at short times. This had been clearly manifested by a linear stability analysis of the
equation for the first moment, and was understood as a renormalization
of the linear forces due to the noise term. Let us
now perform in our model a similar linear stability analysis of the
homogeneous disordered state $\phi=0$
\cite{becker94,Ibaneslin} and we will be able to show that this is not
the mechanism that
induces the phase transition in our model. The dynamical equation for
the first statistical moment, in the linear approximation, reads
\be
\frac{\partial \langle\phi\rangle}{\partial t}
=-(a+\varepsilon_0\,c)\langle
\phi\rangle + K \nabla^2\langle\phi\rangle\,.
\ee
where Novikov's theorem has been applied to the term 
$\langle \Gamma(\phi)^{1/2}\xi\rangle$. 

The contribution of the noise is in the same direction as the one  from the
local potential, and thus the homogeneous state $\phi=0$ is  linearly stable. 
Hence the physical mechanism responsible for the appearance of a new phase must
be different from the one of previously known cases of noise induced
phase transitions. A simple interpretation can be found in terms of a balance
between the role of the deterministic monostable local potential, which tends
to take the system towards the disordered ($\phi=0$) phase, and the stochastic
motion, which is due to the fact that fluctuations are more intense in the
disordered phase, and thus push the system away from it. The presence of a
spatial coupling $K$ helps to break the symmetry of the homogeneous state. 

\begin{figure}[h]
\begin{center}
\epsfig{file=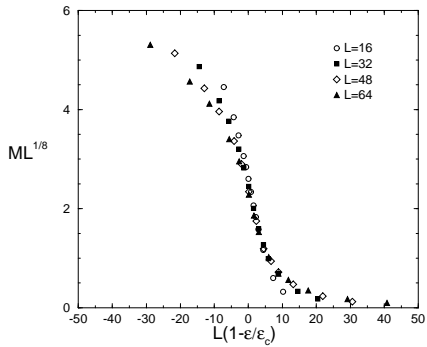, width=6.0cm}
\end{center}
\caption{\label{fig:3} Finite--size scaling function of the order 
parameter $M$ with the 2d-Ising critical exponents.} 
\end{figure}

\begin{figure}[h]
\begin{center}
\epsfig{file=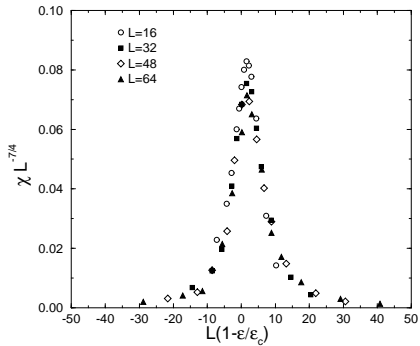, width=6.0cm}
\end{center}
\caption{\label{fig:4} Finite--size scaling function of the susceptibility
$\chi$ with the 2d-Ising critical exponents.}
\end{figure}

A natural question arises concerning the universality class of the
noise-induced phase transition described above. The availability of
an exact nonequilibrium potential for this model, given by Eq.
(\ref{freeenergy}), allows us in a very simple way to analyze its
universal behavior at the critical point. If an expansion of
$\,\ln\Gamma(\phi)$, as chosen in Eq. (\ref{eq:M}), is performed on
(\ref{freeenergy}), the nonequilibrium effective potential can be seen 
to have a Ginzburg-Landau form with infinite higher-order (irrelevant)
terms. Since the universality class of the Ginzburg-Landau model is
that of the Ising model, we predict that this transition will belong
to the d-dimensional Ising universality class. 
This is indeed confirmed by a finite-size scaling analysis of the
numerical results. We make use of the following scaling laws,
generalized from equilibrium statistical mechanics \cite{nises}:
\begin{eqnarray}
\label{fsscalm}
M & = & L^{-\beta/\nu}\,\widetilde M\,(\tilde \varepsilon\, L^{1/\nu})\,,
\\
\label{fsscalx}
\chi & = & L^{\gamma/\nu}\,\widetilde \chi\,(\tilde \varepsilon\, L^{1/\nu})\,
\end{eqnarray}
where $\widetilde M$ and $\widetilde \chi$ are the corresponding scaling
functions, and $\tilde \varepsilon$ is the reduced control parameter,
$\tilde \varepsilon=(1-\varepsilon/\varepsilon_c)$. The values of the
critical exponents $\beta$, $\gamma$ and $\nu$ define the
universality class. We have chosen the corresponding values of the
2-d Ising universality class, and plotted expressions (\ref{fsscalm})
and (\ref{fsscalx}) in Figs. \ref{fig:3} and \ref{fig:4}.
All data are seen to scale fairly well into the same curve in the region
close to the critical point. As expected, those data corresponding to 
the smallest system size ($L=16$) show the major discrepancies.  
Hence, we can conclude that this new kind of noise-induced phase transitions 
belongs to the 2d-Ising universality class.

In conclusion, we have introduced a dynamical system that
exhibits a nonequilibrium phase transition from disorder to order as the
intensity of the noise is increased. As a consequence, the system is disordered
below a certain amount of noise and can be considered a possible scenario,
within the context of stochastic differential equations, for the appearance of a
lower critical temperature in physical systems. At variance with other
noise-induced nonequilibrium phase transitions reported before, the physical
mechanism inducing the transition is not a noise--induced short time
instability of the disordered state. Rather, we are able to show that the
effective local potential, which can be obtained analytically, exhibits
bistability as the intensity of the noise increases, whereas the deterministic
one is monostable. Hence, the mechanism is seen to be given by a balance
between deterministic forces and stochastic motion. Our model shows that it is
possible to obtain a noise--induced {\sl phase} transition by coupling
adequately 0--dimensional systems undergoing a noise--induced transition.
We have argued that this type of coupling is consistent with a mesoscopic
description in terms of fields, kinetic coefficients and the
fluctuation--dissipation relation. Our
findings have been sustained by numerical simulations in a two-dimensional
lattice model. Finally, by studying the exact stationary probability
distribution, we are able to show that the model belongs to the same
universality class as the Ising model.

We acknowledge the Direcci\'on General de Ense\~nanza Superior (Spain)
for financial supports under projects PB94-1167, PB96-0241, PB97-0141-C02-01, and
PB98-0935.


\end{multicols}


\begin{references}
\bibitem{cutoff} As it is well known, calculations in a continuum space
in the presence of a multiplicative white noise have to be considered
with care \protect\cite{nises,MT97}. Therefore, we interpret our noise
here to have a cutoff at short length scales, according to standard
approaches in nonequilibrium critical dynamics. Using, for instance, a 
regular lattice of spacing $\Delta x$ for regularization, we find
$\varepsilon = \varepsilon_0/ \Delta x^d$. 
More details on our
calculations will be published elsewhere.
\end{references}
\end{document}